# The Bose-glass phase in twinned $YBa_2Cu_3O_{7-\delta}$

S. A. Grigera, E. Morré, E. Osquiguil, C. Balseiro, G. Nieva and F. de la Cruz
*Centro Atómico Bariloche and Instituto Balseiro,*
*Comisión Nacional de Energía Atómica, 8400 San Carlos de Bariloche, Argentina*

Using an extensive scaling analysis of the transport properties in twinned $YBa_2Cu_3O_{7-\delta}$ crystals we have experimentally found the predicted change in the universality class of the Bose-glass to liquid transition when the magnetic field is applied at small angles away from the direction of the correlated defects. The new dynamical critical exponent is $s' = 1.1 \pm 0.2$.

Since the pioneering work by Gammel et al. [1] showing the existence of a solid-liquid transition in the vortex lattice, much progress has been made in the understanding of the $H - T$ phase diagram of high temperature superconductors. New theoretical concepts describing the vortex structure as a distinctive form of condensed matter have been introduced to understand its transport and thermodynamic properties. In particular, the so-called "irreversibility" line was interpreted as a second order thermodynamic transition from a vortex liquid to a superconducting vortex-glass [2,3] phase, in which the topological disorder induced by point defects plays a major role.

The analysis of the current-voltage (I-V) characteristics measured [4,5] in twinned $YBa_2Cu_3O_{7-\delta}$ single crystals with the magnetic field applied in the $c$-direction, in terms of the vortex-glass scaling theory [3], provided conclusive evidence of a genuine second order phase transition. Despite this, questions on the existence of a vortex-glass phase have been raised [6] based on the preeminence of correlated disorder in the measured samples. The observation of a cusp in the "irreversibility line" [7] for fields close to the $c$-axis in twinned $YBa_2Cu_3O_{7-\delta}$ crystals, together with transport measurements showing a sharp increase in the resistivity as the magnetic field was rotated away from the twin boundary orientation [8] led Nelson and Vinokur to introduce a new interpretation of the second order phase transition in the vortex lattice in the presence of correlated defects. [6,9] This new low temperature glassy phase stabilized by correlated defects is known as the Bose-glass.

Recent experiments [10,11] in samples where point defects dominate the transport properties cast doubts about the existence of the vortex-glass phase because the suppression of the first order melting does not lead to a second order transition but rather to a continuous freezing of the vortex structure.

For fields parallel to the $c$-axis, the Bose-glass theory [9] predicts for the (I-V) characteristics similar critical exponent relations as those given by the vortex-glass. The main difference between both theories arise when the magnetic field is rotated off the $c$-axis. The Bose-glass theory predicts a critical state with a different universality class when the field is tilted away from the correlated defects compared to that for perfect alignment. This change in the universality class reflects the different mechanisms by which the Bose-glass phase is melted into a liquid when the magnetic field is rotated off the correlated defect direction. [12] Therefore, in order to differentiate one glass from the other, the response of the transport properties to magnetic field tilting in the critical region must be investigated.

In this paper we present measurements of the electrical transport properties near the solid liquid transition in twinned $YBa_2Cu_3O_{7-\delta}$ crystals as a function of the temperature and the angle between the applied magnetic field and the twin boundaries. The resistivity shows a pronounced change in its temperature dependence as the field is tilted an angle $\theta$ away from the $c$-axis, while the transition temperature shows a sharp cusp around $\theta = 0$. A scaling analysis of these results in terms of the Bose-glass theory, applied to $\theta \neq 0$, accounts for the observed features in the resistivity and transition temperature. Furthermore, the experiments allow the determination of the new dynamical critical exponent that characterizes the universality class of the transition when the field is tilted away from the defects. This demonstrate that the second order transition previously observed [4] in this type of crystals when the field is applied in the $c$-direction does not correspond to a vortex-glass transition but to a Bose-glass one.

The heavily twinned $YBa_2Cu_3O_{7-\delta}$ single crystal was grown as indicated in Ref. [13], had a critical transition temperature $T_c \approx 93K$ and a transition width $\Delta T_c \leq 0.3K$ with dimensions $(1 \times 0.5 \times 0.03)mm^3$ with a single family of twin planes parallel to its length. The crystal was mounted onto a rotatable sample holder with an angular resolution of $0.06°$ inside a cryostat with an 8T magnet. The $ab$ voltages were measured as a function of temperature using a lock-in amplifier at 29 Hz. For the resistivity measurements current densities $J \leq 5A/cm^2$ were applied through the crystal along the twin planes. The rotation of the magnetic field off the $c$-axis was done at constant Lorentz force for fields between 1T and 8T, similar results were obtained for all magnetic fields. The data shown in this paper are for a field of 6 T.

In the Bose-glass transition [9], like in any other continuous transition [14], the properties of the system in the region of the phase diagram surrounding the phase boundary are dominated by fluctuations. In this zone, called *critical region*, the dynamical response of the vortex lattice can be obtained via a scaling theory by equating the appropriate dimensionless quantities [3].

In particular the relation between the current density $J$ and electric field $E$ when the magnetic field is aligned with the correlated defects obeys [9]

$$E \mid t \mid^{-\nu(z+1)} = F_{\pm}(\mid t \mid^{-3\nu} J\phi_0/c), \tag{1}$$

where $\nu$ and $z$ are the exponents governing the size and time relaxation of fluctuations respectively, $t = (T - T_{BG})/T_{BG}$, $T_{BG}$ is the Bose glass transition temperature and $F_{\pm}$ are two analytic functions for $t > 0$ and $t < 0$ respectively.

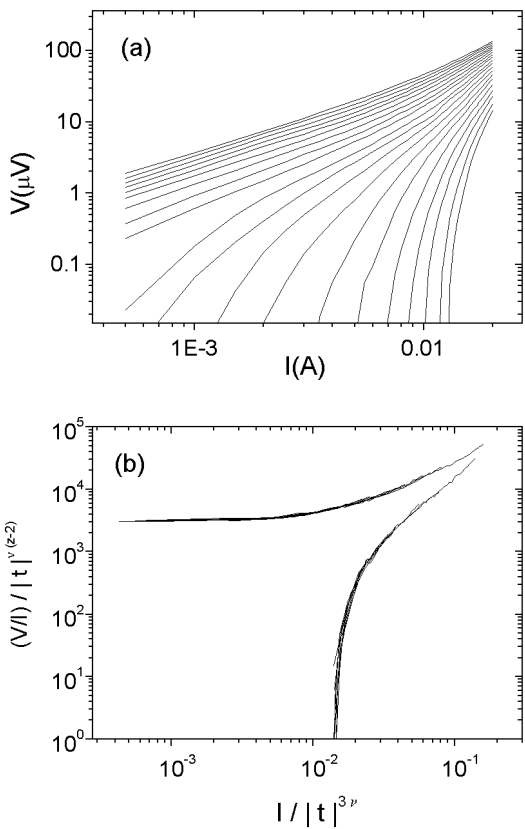

FIG. 1. (a) Current voltage characteristics between temperatures 83.5 and 81.4 K at intervals of 0.1 K from top to bottom for $H \parallel c$. (b) Scaling of the I-V curves in (a) according to equation 1

This universal behavior for the I-V characteristics has been thoroughly investigated [4,5,15]. In fig.1 we show the corresponding scaling for our I-V measurements for different temperatures within the critical region. The exponents thus obtained are $\nu = 1 \pm 0.2$ and $z = 6 \pm 1$, and the glass temperature is $T_{BG} = 82.6 \pm 0.05 K$.

Equation 1 can be extended to the case where the field is applied at small angles $\theta$ with respect to the correlated defects. In the liquid phase and in the small $J$ limit (linear regime) the resulting equation can be linearized into [9]

$$\rho(t, \theta) \approx \mid t \mid^{\nu(z-2)} f_{\pm}(x), \tag{2}$$

where $t = (T - T_{BG})/T_{BG}$, $T_{BG}$ is the Bose Glass temperature at $\theta = 0$, and $x = \sin(\theta) \mid t \mid^{-3\nu}$. The two subscripts $+/-$ in the scaling function refer to $t > 0$ and $t < 0$ respectively.

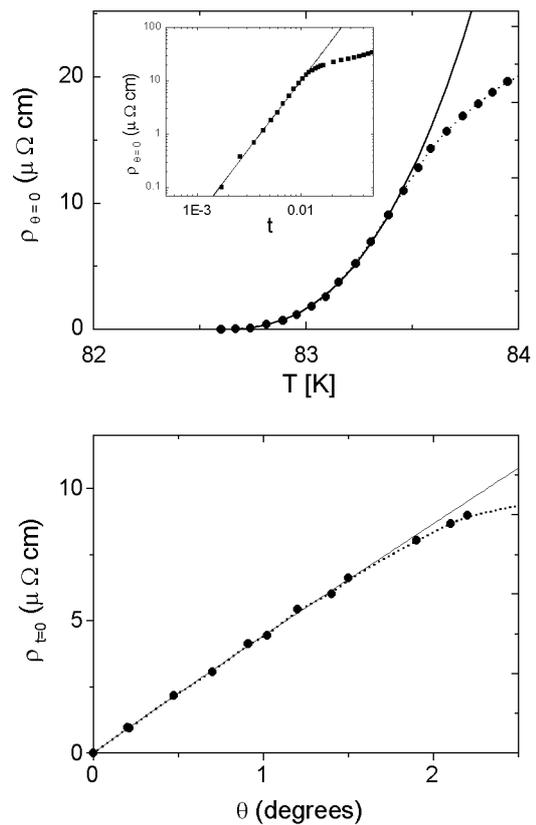

FIG. 2. Upper panel: Temperature dependence of the resistivity ($\rho$) for $\theta = 0$ (solid symbols) and fit using $\rho \sim t^s$ (solid line) Inset: logarithmic plot of the same data versus the reduced temperature $t$. Lower panel: $\rho$ at $T = T_{BG}(0)$ as a function of $\theta$ (solid symbols) and fit according to equation 5 (solid line).

The function $f_+$ is analytic and even, but to describe the transition to the Bose-glass phase for finite angles $f_-$ should have [9] a singularity at $x = \pm x_c$. If we call $s'$ the exponent with which the resistivity vanishes in this case (*i.e.* $\rho \sim \mid T - T_{BG}(\theta) \mid^{s'}$) then $f_-$ should behave as [16]

$$f_-(x) \sim (\mid x \mid -x_c)^{s'}, \tag{3}$$

for $|x| > x_c$ and $f_-(x) = 0$ otherwise. The critical temperature as a function of the angle is given by

$$T_{BG}(\theta) = T_{BG}(0)\left[1 - \left(\frac{|\sin\theta|}{x_c}\right)^{1/3\nu}\right]. \quad (4)$$

Let us first analyze two particular cases of equation 2, $\theta = 0$ and $t = 0$. In the first case the vanishing of the resistivity is governed by a power law of the form $\rho \sim t^s$ where $s = \nu(z-2)$. Indeed the experimental data for the temperature dependence of the resistivity for $\theta = 0$ shown in Fig. 2 is fitted (solid line) by a single power law with $s = 2.8 \pm 0.2$ and $T_{BG} = 82.6K$ which are in agreement with the values of $z$, $\nu$ and $T_{BG}$ found from the scaling of the I-V curves.

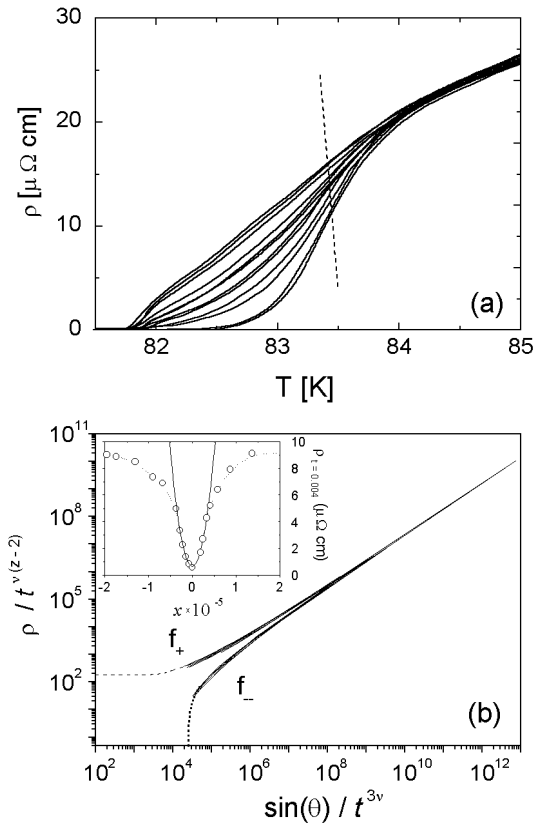

FIG. 3. (a) Resistivity versus temperature for different tilting angles. From top to bottom: $\theta$ = 2.1, -2.1, -1.9, 1.4, -1.4, -1.17, 0.92, -0.91, -0.7, 0.47, 0.2, -0.2. The dashed line indicates the upper limit of the critical region (see text) (b) Scaling of the curves shown in (a) according to eq. 2 (solid lines), the dotted line is a fit according to eq. 3, the dashed line is the behaviour of $f_+$ for $x \to 0$ obtained from the fit shown in the inset, where the angular dependence of the resistivity at a temperature slightly above $T_{BG}$ is shown.

In the particular path of the phase diagram determined by $t = 0$, the resistivity is given by [9]

$$\rho(0,\theta) \approx |\theta|^{(z-2)/3}, \quad (5)$$

for small $\theta$. This relation is verified experimentally as shown in the inset of Fig. 2, where the fit (solid line) gives $z = 5.5 \pm 1$ in agreement with the previous value obtained for $z$.

The resistivity in the linear regime as a function of temperature for different angles, see Fig. 3a, should collapse into two single curves when plotted using the rescaled axes according to equation 2. Indeed, the data show an excellent collapse with $\nu = 1 \pm 0.2$ and $z = 6 \pm 1$ as depicted in Fig. 3b. The agreement with the values obtained from the I-V scaling for $\theta = 0$ is remarkable.

The fit (dotted line) of the experimental data of the lower branch ($f_-$) for small $x$ in Fig. 3b using expression 3 with $x_c$ and $s'$ left as free parameters, gives $x_c = 25000 \pm 3000$ and $s' = 1.1 \pm 0.3$. Note that the new critical exponent $s'$ is considerably smaller than that corresponding to $\theta = 0$.

The inset of Fig.3b shows the angular dependence of the resistivity at a temperature close to $T_{BG}(t = 4.810^{-3})$ as a function of $x$. The solid line is a fit to the data using a quadratic approximation to $f_+(x)$, that is [9], $\rho(t,\theta) \approx \rho_0 |t|^s [1 + Ax^2]$. This fit allows an extrapolation of the function $f_+$ to the limit $x \to 0$ as plotted with a dashed line in Fig.3b.

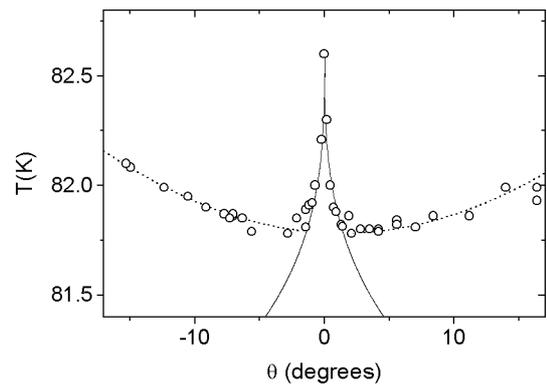

FIG. 4. Angular dependence of T($\rho = 0$) (open symbols). The solid line is the Bose-glass transition temperature according to equation 4 with the parameters $x_c = 25000$ and $\nu = 1$ obtained from the fit to $f_-$ in Fig. 3. The dashed line is the expected variation of $T(\rho = 0)$ due to the anisotropy.

The analysis of the range of temperatures and angles within which the experimental data collapse into the universal curves gives the upper bounds of the critical region. The limit in temperature is indicated in Fig. 3 by a dashed line. Curves for $\theta > 2.1°$ do not show critical behavior for any temperature.

The different universality class for $\theta \neq 0$ gives a natural explanation for the experimentally detected change in the temperature dependence of the resistivity when approaching $\rho = 0$ for magnetic fields off the c-axis (see Fig. 3 and Ref. [18,19] for data at 6T and Fig 16 in Ref. [8] for data at 2T).

As discussed before, the singular point $x_c$ and the exponent $\nu$ fix the phase boundary $T - \theta$. Using the $x_c = 25000$ obtained from the scaling of the $\rho(t,\theta)$ and $\nu = 1$ in expression 4 the full lines in Fig 4 are obtained. In the same figure the experimental Bose-glass transition temperatures (defined as $T(\rho = 0)$ in the linear regime within the experimental resolution of $1\mu\Omega$) are plotted (open dots). Notice the excellent agreement between the experimental data and the theoretical curve obtained without using free parameters.

It is interesting to note that the angle above which the points begin to follow the intrinsic anisotropy of the material (dotted line) coincides with the angle $\theta = 2.1°$ above which the curves do not follow critical scaling laws. This is in agreement with the crossover to a first order phase transition line as previously suggested [17].

In conclusion we have shown via an extensive scaling analysis of the transport properties in twinned $YBa_2Cu_3O_{7-\delta}$ crystals in the neighborhood of the "irreversibility line" that this line corresponds to a Bose-glass transition, and therefore no experimental evidence for the existence of the vortex-glass phase has been provided until now. We have experimentally found the predicted critical crossover when the magnetic field is applied at small angles away from the correlated defects and determined the new dynamical exponent $s' = 1.1 \pm 0.2$ governing the vanishing of the resistivity for angles different from zero and temperatures close to the transition.

These results encourage experimental studies of the behavior of the thermodynamic properties near the Bose-glass transition, in particular the transverse Meissner effect [9,20].

We acknowledge illuminating discussions with David R. Nelson. This work was partially supported by CONICET PIP 4207 and Fundación Antorchas under grant A13359/1-000013. S. A. G. holds a fellowship from the Consejo Nacional de Investigaciones Científicas y Técnicas (CONICET), E.M. holds a fellowship of the Gottlieb Daimler-und-Karl Benz-Stiftung. E. O., C. B. and G.N. are members of CONICET.